\documentclass{elsart5p}

\usepackage{graphicx}

\usepackage{amssymb}

\begin{document}

\begin{frontmatter}



\title{The Puzzle of Neutron Lifetime}


\author{Stephan Paul}

\address{
Physics Department E18 and Cluster of Excellence Exc153\\
Technische Universit\"at M\"unchen \\
James Frank Str.\\
D-85748 Garching}

\begin{abstract} In this paper we review the role of the neutron lifetime and discuss the present status
of measurements. In view of the large discrepancy observed by the two most precise individual
measurements so far we describe the different techniques and point out principle strengths and weaknesses. In
particular we discuss the estimation of systematic uncertainties and its correlation to the statistical
ones. In order to solve the present puzzle, many new experiments are either ongoing or being proposed. An
overview on their possible contribution to this field will be given.

\end{abstract}


\end{frontmatter}

\section{Introduction} The lifetime of the free neutron is a basic physical quantity, which is relevant in a variety of different fields of particle and astrophysics. Being directly related to the weak interaction characteristics it plays a vital role in the determination of the basic parameters like coupling constants or quark mixing angles as well as for all cross sections related to weak $p-n$ interaction. We shall briefly give an overview on such processes:
\subsection{Astrophysics} 
\begin{figure}\begin{centering}
         \includegraphics[width=7.5cm]{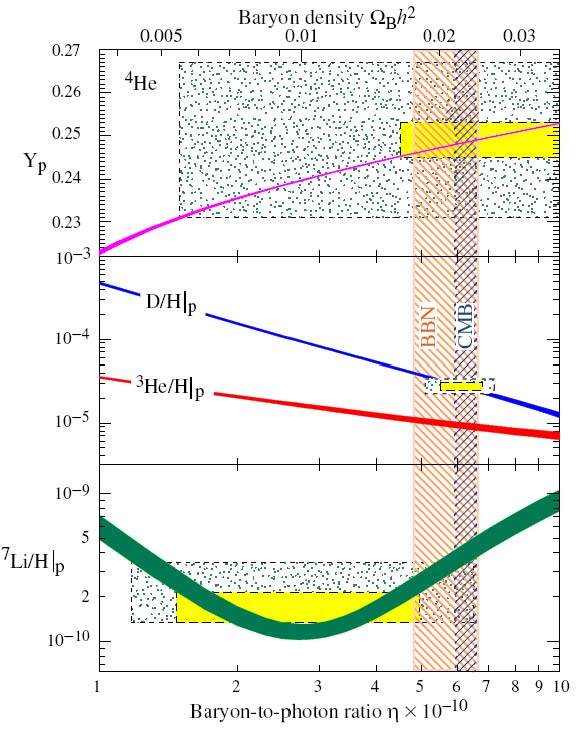}\\
         \caption{Isotope mass fraction versus the baryon fraction in the universe using the standard model of cosmology (lines). The solid areas depict the statistical errors for astronomical observations, the dotted ones the systematic uncertainties \cite{BBNS}.}\label{helium_abundance}\end{centering}
      \end{figure}

\begin{figure}\begin{centering}
         \includegraphics[width=5cm]{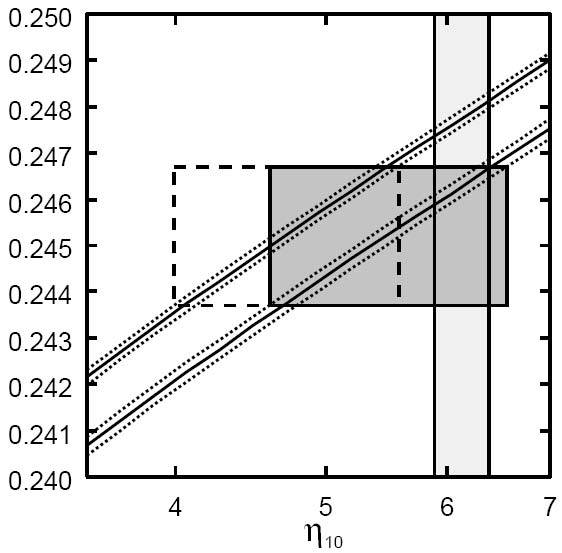}\\
         \caption{Helium mass fraction versus the baryon fraction in the early universe for two different values of the neutron lifetime \cite{BBNS_2}. Upper: PDG value \cite{PDG06}, lower: ref. \cite{serebrov_1}. The boxes show the allowed values for the baryon mass fraction with the box size indicating statistical uncertainties for $Y_p$ only. The vertical band shows the baryon fraction deduced from WMAP data.
}\label{helium_abundance_lifetime}\end{centering}
      \end{figure}

One of the key processes with relevance to neutron decay is primordial nucleosynthesis \cite{BBNS}. A few minutes after the big bang weak interaction causes an almost equilibrium of neutrons and protons owing to the reactions $n\rightarrow p e^-\overline{\nu}_e$ and the electron capture reactions $p e^-\leftrightarrow n\nu_e$ and $n e^+\leftrightarrow p\overline{\nu}_e$. The equilibrium of these reactions is broken once the expansion rate of the universe wins over the mean free path of the neutrinos (governed by the strength of the weak interaction $\Gamma_{n\leftrightarrow p} \sim G^2_{\mathrm{F}}\cdot T^5)$. At this temperature $T$ neutrinos decouple from the system and $T$ determines the $n/p$ ratio $n/p = e^{-Q/T}$,
where $Q=1.293~\rm{MeV}$ is the neutron-proton mass difference. This ratio changes subsequently owing to free neutron decay. As the universe expands the temperature drops below the photo-dissociation threshold for deuterons and efficient nucleosynthesis starts, leading to the production of light elements like  deuterium, helium and lithium. The abundance predictions of the standard model of cosmology using the neutron lifetime as input parameter is shown in fig. \ref{helium_abundance} as function of the
 baryon-to-photon ratio $\eta_{10}$, where $Y_{\mathrm{P}}$ denotes the helium mass
 fraction in the early universe \cite{BBNS}.
 Figure \ref{helium_abundance_lifetime} demonstrates, as an example, the effect of changing the neutron lifetime in the model \cite{BBNS_2}. Although
 having big influence,
 the value of $Y_{\mathrm{P}}$ determined from low metalicity regions is not yet measured with good enough precision and systematic uncertainties in the extrapolation of the helium abundance to regions with zero metalicity dominate the
 experimental error band. Thus, the consistency of the standard model is not in question.\\


\subsection{Particle Physics} In the standard model, neutron decay is governed by weak interaction with
the underlying V--A structure. The lagrangian contains two parts, a leptonic and a hadronic one. The latter one is written as \cite{decay_formula}:
\begin{eqnarray}
\label{V_A}
    V_\mu-A_\mu & = & \emph{i}\overline{\Psi}_p\{f_1(q^2)\gamma_{\mu}+f_2(q^2)
    \frac{\sigma_{\mu\nu}q^{\nu}}{m_p}+\textit{i}f_3(q^2)\frac{q_{\mu}}{m_e}\}\Psi_n\nonumber\\
    && -\emph{i}\overline{\Psi}_p\{f_i\rightarrow g_i\gamma_5\}\Psi_n.
\end{eqnarray}
Using the conserved vector current (CVC) hypothesis most form factors $f_i$ and $g_i$ can be set to zero but
\begin{eqnarray}
\label{V_A_2}
    G_{\mathrm{V}}=f_1(q^2\rightarrow 0)\cdot V_{ud}\cdot G_{\mathrm{F}} = g_{\mathrm{V}}\cdot V_{ud}\cdot G_{\mathrm{F}}\nonumber\\
    G_{\mathrm{A}}=g_1(q^2\rightarrow 0)\cdot V_{ud}\cdot G_{\mathrm{F}} = g_{\mathrm{A}}\cdot V_{ud}\cdot G_{\mathrm{F}}\nonumber
\end{eqnarray}
and we obtain an expression for the first element of the Cabibbo-Kobayashi-Maskawa quark mixing matrix $V_{ud}$
\begin{equation}
\label{V_A_3}
    {\rm with}~~\lambda=\frac{}{}\frac{G_{\mathrm{A}}}{G_{\mathrm{V}}}=\frac{g_{\mathrm{A}}}{g_{\mathrm{V}}};~~~~~
    \mid V_{ud}\mid^2=\frac{1}{\tau_n}\frac{(4908.7\pm1.9)~{\rm s}}{(1+3\lambda^2)}.\nonumber
\end{equation}

Largest theoretical uncertainties come from radiative corrections which are common to both, free neutron decay and pure Fermi-transitions in nuclei \cite{hardy08}.
\subsection{Exotic implications}
The energy production in our sun proceeds predominantly via two processes, ${pp}$ fusion and the CNO cycle \cite{adelberger}. The relative strength of the two processes depends among other on the coupling strength determining the ${pp}$ fusion which in turn involves $G_{\mathrm{A}}$. Thus, the neutrino spectrum from the sun depends indirectly on the neutron lifetime and its uncertainty. However, the temperature ($T$) dependence of other processes is enormous and masking this effect (e.g. the $^8$B rate is proportional to $T^{25}$).\\
On the other hand neutrino cross sections relevant in all neutrino experiments \cite{declais} also involve $G_{\mathrm{A}}$. In turn the measured neutrino cross-section directly yields the neutrino helicity $\textsl{H}_{\nu}$ and thus $G_{\mathrm{A}}$ is linked to a fundamental neutrino property in weak interaction \cite{neutrino_crosssection}:
\begin{displaymath}
R=\frac{\sigma(\overline{\nu}+p\rightarrow n+e^+)}{\sigma_{\mathrm{expected}}}= \frac{1}{2}(1+\textsl{H}_{\overline{\nu}}).
\end{displaymath}
Here $\textsl{H}_{\overline{\nu}}$ is the anti-neutrino helicity. A lower value for the neutron lifetime (as inferred from more recent measurerment - see next secion) would result in  higher coupling constants and thus the cross section and in turn raises the lower limit for $\textsl{H}_{\overline{\nu}}$ to $\textsl{H}_{\overline{\nu}}>0.97$ (assuming $\tau_n=878~\rm s$).
\section{Measuring methods for the Neutron Lifetime}
Two general methods exists to determine $\tau_n$: \textit{in beam} and \textit{storage} experiments. In the first method a neutron beam passes a fiducial decay volume and the number of decay products is recorded. Absolute count rates are needed for the neutron flux and the number of decay particles as well as a precise and stable knowledge of the decay volume. Uncertainties due to spectral effects in the neutron velocity distribution cancel to first order as the neutron detection efficiency and the effective exposure time have the same velocity dependence $\varepsilon\sim 1/{v_n}$ (for the common case of small n-detection efficiencies).
For the latter group of experiments only relative count rates are important (measurement of the exponential shape of the decay-time distribution) but the measured lifetime always is a combination of two effects, the $\beta$-decay rate and a loss rate which can have various origins but typically exhibits strong spectral dependence.
\begin{equation}
\label{lifetime_storage}
    \frac{1}{\tau_n}= \frac{1}{\tau_{\beta}} + \frac{1}{\tau_{\mathrm{loss}}}.
\end{equation}
Thus, both methods are complementary in their systematic uncertainties.
\subsection{In-beam measurements}
\begin{figure}\begin{centering}
   \includegraphics[width=9cm]{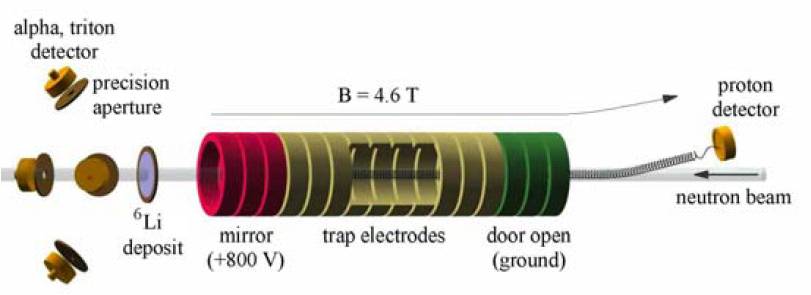}\\
   \includegraphics[width=6cm]{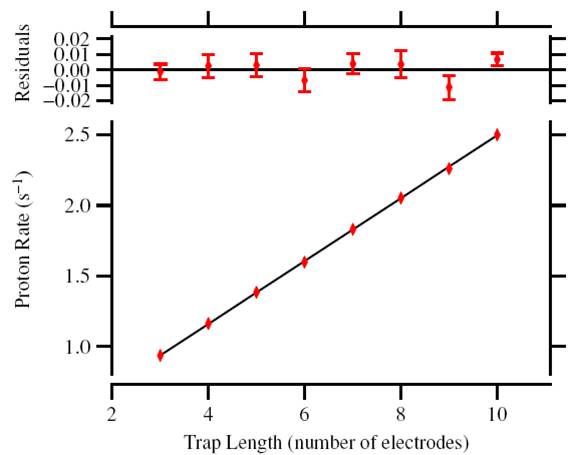}\\
    \caption{\textit{In-beam} experiment with proton counting at NIST. Upper: Experimental setup with Penning trap, proton counters and neutron flux measurement. Lower: Proton count rate for different lengths of the Penning trap (see eq. \ref{proton_counting}) \cite{nico}.}\label{lifetime_proton_decay_NIST}\end{centering}
  \end{figure}

\textit{In-beam} measurements have the longest traditions and were the base for the first
determination of the neutron lifetime. Robson \cite{robson} in his experiment
extracted protons from a fiducial decay volume and estimated the neutron lifetime to be between 9 and 25 min. The latest of such experiments obtained a more than 100 times higher precision. Key ingredients to this experiment are a well controlled fiducial decay volume, which is made from a set of ring shaped electrical cathodes which define a trapping volume (Penning trap) for decay protons (see fig. \ref{lifetime_proton_decay_NIST}) and very well calibrated particle detectors. Accumulated over a preset time decay protons are extracted onto a proton counter. The decay volume can be extended by equal portions \textit{$n\cdot L$} ($L$ being the length of a trap subsection) and the lifetime extracted according to the differential count rates
\begin{equation}
  \frac{N_{\mathrm{proton}}}{N_{\mathrm{neutron}}}=\tau^{-1}_{n}\cdot \frac{\epsilon_{\mathrm{proton}}}{\epsilon_{\mathrm{neutron}}}\cdot (n\cdot L+L_n).
  \label{proton_counting}
  \end{equation}

Well measured efficiencies for proton and neutron counting lead to corrections to $\tau_{n}$ of -5.3~s and +5.2~s, respectively, thus almost cancel each other and the overall correction applied to the measured value is -0.4~s leading to the most precise measurement within this class of experiments of $\tau_{n}=886.3\pm 1.2_{\mathrm{stat}}\pm 3.2_{\mathrm{syst}}~{\rm s}$.
\subsection{Stored neutrons}
The second and very successful method to determine $\tau_{n}$ is based on the storage of neutrons in a bottle made from magnetic fields or suitable wall materials. At any rate, so far the lifetime has been extracted by counting survival neutrons after a well defined holding time $T_{\mathrm{store}}$. Varying $T_{\mathrm{store}}$ the lifetime can be determined. A more efficient but also more elaborate technique is the real time detection of decay particles from the stored neutrons which has only been used once so far (see section \ref{NIST}) and will be employed by a forthcoming experiment (see section \ref{future_measurements}).
\subsubsection{Magnetic storage}\label{NIST}
Magnetic gradient fields provide a potential for magnetic moments causing a force
 \begin{displaymath}
 \overrightarrow{F}=-\overrightarrow{\nabla}(\overrightarrow{\mu}\cdot \overrightarrow{B})
  \end{displaymath}
where
  \begin{displaymath}
  \mu_n=-60.3~{\rm neV/T}~; ~~~B_{\mathrm{max}}^{\mathrm{typ}}\sim 2-3~{\rm T}.
 \end{displaymath}
 Such bottles can use a combination of fields to obtain a closed system:
\begin{itemize}
  \item magnetic fields and centrifugal forces (NESTOR storage ring)
  \item magnetic fields and gravitational forces (magneto-gravitational trap)
  \item $4\pi$ magnetic containment (Ioffe trap)\\
\end{itemize}

\par
\noindent
This variety shall be described in the following:

\par
\noindent
\begin{description}
  \item  \textit{Neutron storage ring:} In the late seventies the first storage ring for neutral particles was designed and constructed (see fig. \ref{nestor} \cite{trinks77}). Using a magnetic sextupole field neutrons traveling along circular orbits are experiencing restoring forces, which are proportional to their radial distance from the nominal orbit. Coils mounted outside and on top and bottom of the circular storage volume provide fields such that $F\sim \triangle r, \triangle z$, respectively. Restoring forces for neutrons traveling at too small orbits are provided by the centrifugal potential. This scheme resembles a betatron and thus neutron orbits will undergo betatron oscillations, the amplitude of which has to be limited by suitable neutron-beam shaping in the initial stage of storage.\\

  \begin{figure}[h]\begin{centering}
   \includegraphics[width=7cm]{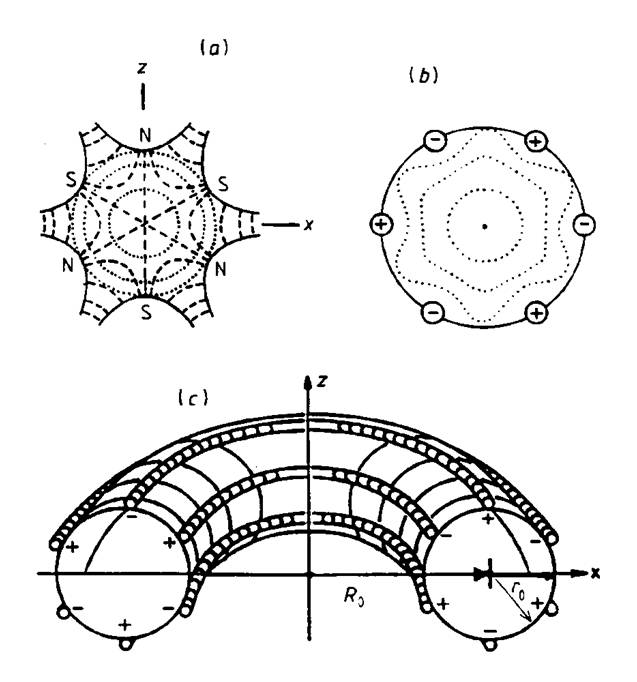}\\
   \caption{Conceptual drawing of the neutron storage ring for very-cold neutrons. a) ideal sextupole with field lines and equipotential lines, b) single-wire realization of a sextupole, c) storage-ring realization of a sextupole \cite{paul}.}\label{nestor}\end{centering}
  \end{figure}

      This apparatus was placed at the ILL turbine and neutrons with tangential velocities $v_n^{\|}\sim 50~{\rm m/s}$ were stored and counted after a preset holding time. The lifetime was measured to $\tau_{n}=877\pm 10_{\mathrm{stat}}~{\rm s}$ (see fig. \ref{nestor_lifetime}) \cite{paul} with no corrections applied and conservative error estimates ($\chi^2/{\mathrm{ndf}}\approx 0.5$). No buildup of betatron oscillations was observed making possible neutron storage time constants up to 3600s.\\

      \begin{figure}[h]\begin{centering}
   \includegraphics[width=7cm]{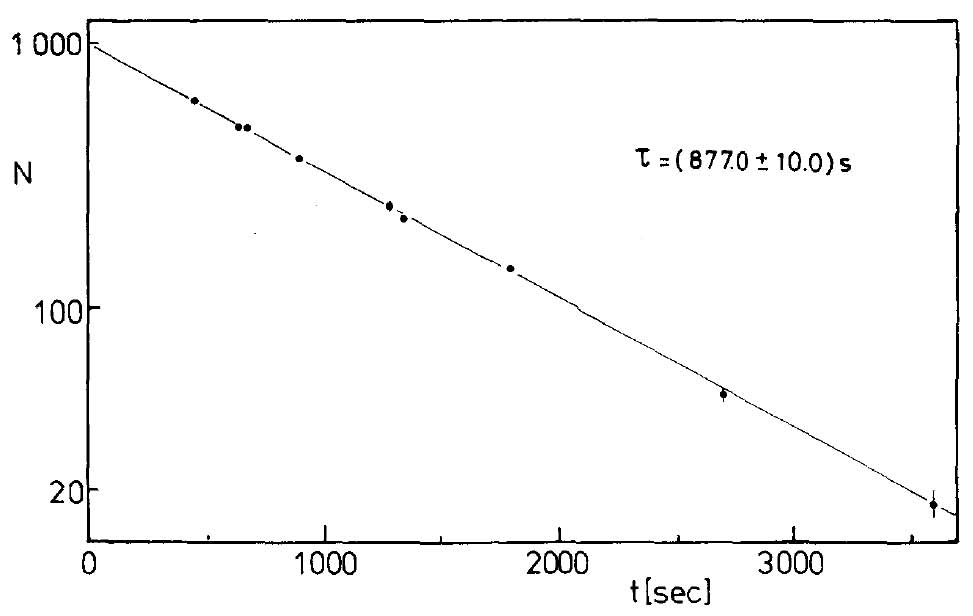}\\
    \caption{Result of the neutron storage experiment including the results of the exponential fit \cite{paul}.}\label{nestor_lifetime}\end{centering}
  \end{figure}

  \item \textit{Ioffe trap:} Ultra-cold neutrons (UCN) with $|v_n|<5~{\rm m/s}$ can be stored in magnetic multipole fields. However, the injection togehter with an efficient source for ultra-cold neutrons is difficult and thus it seems ideal to combine UCN production and storage volume as setup at NIST \cite{IOFFE_1}. \\

  \begin{figure}\begin{centering}
    \includegraphics[width=7cm]{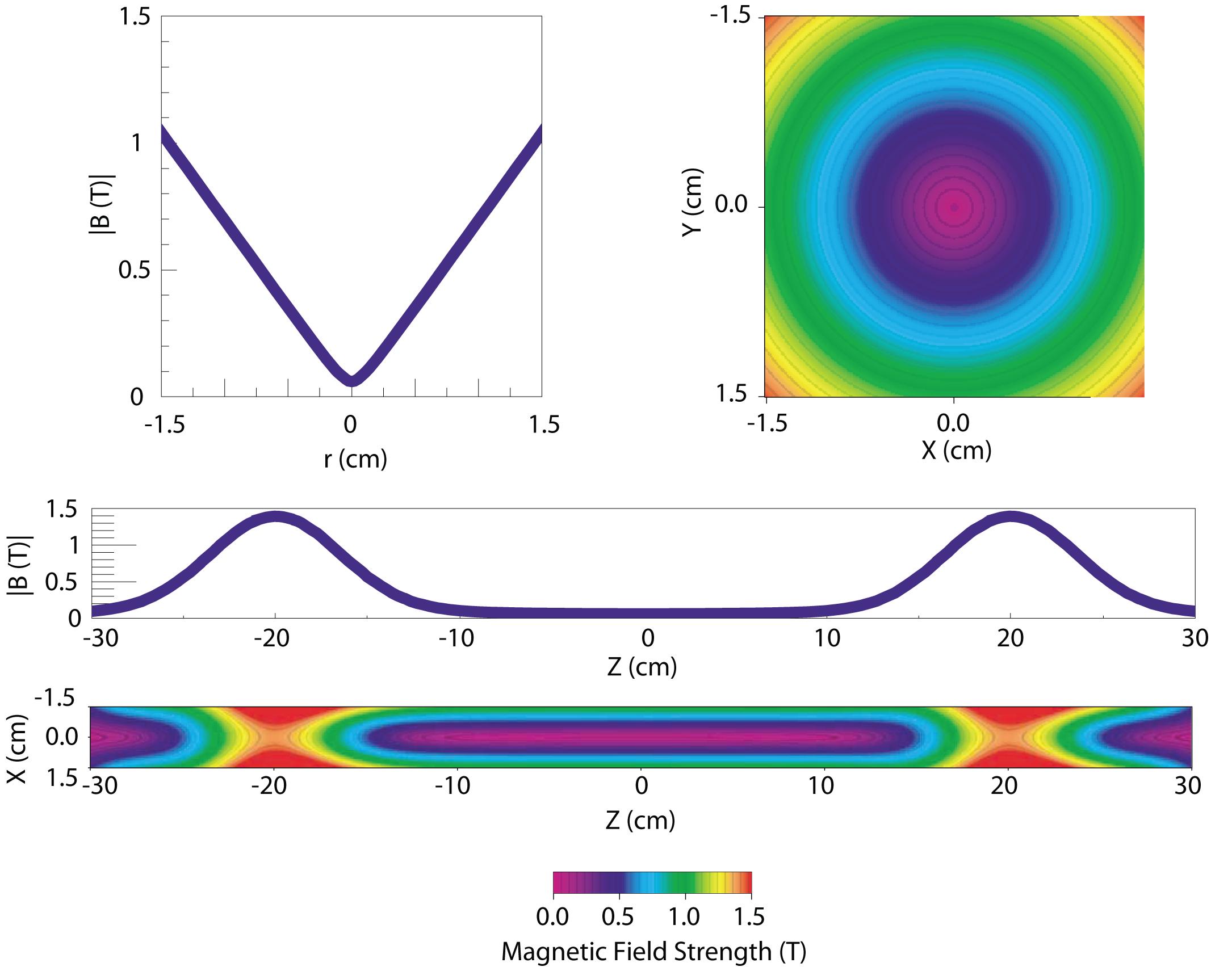}\\
    \caption{Magnetic potential for UCN in the Ioffe-trap: radial dependence (upper), axial dependence (lower) \cite{IOFFE_1}.}\label{IOFFE_fields}\end{centering}
  \end{figure}

    The trap is made from a quadrupole field which is magnetically closed by solenoid-coils at each end of the system (see fig. \ref{IOFFE_fields}).\\

  \begin{figure}\begin{centering}
   \includegraphics[width=7cm]{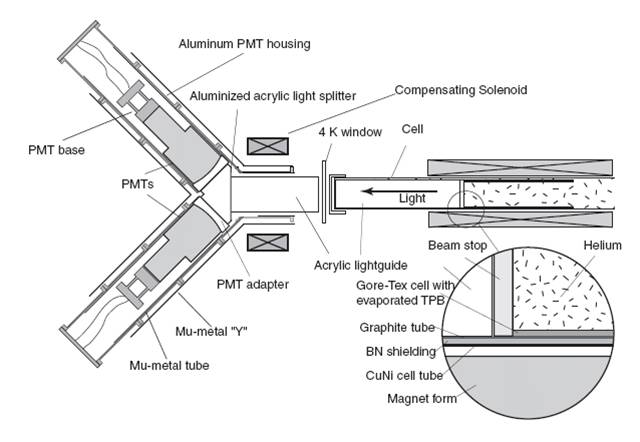}\\
    \caption{Setup of the Ioffe trap for the neutron lifetime experiment. Visible are the UCN converter volume also acting as scintillation detector for electrons. The detection system for the scintillation light is seen on the left.}\label{IOFFE_1}\end{centering}
  \end{figure}

      UCN are produced in superfluid $^4$He and subsequently trapped within the same volume. Neutron decays are observed by detection of decay electrons causing scintillation in the helium. Using reflective painting of the trap walls the scintillation light is funneled to the outside of the trap and detected by two photomultipliers (see fig. \ref{IOFFE_1}). This very complex setup had suffered from background counts caused by the large scintillation volume. In addition, the scheme exhibited a typical problem connected to storage experiments, namely \textit{marginally trapped} neutrons. Neutrons with velocities slightly above the storage potential can be trapped for times $\tau_{\mathrm{loss}}$ if they move on quasi closed orbits with $v_{\bot}<v_{\mathrm{max}}$ but $v_{\mathrm{total}}>v_{\mathrm{max}}$. This effect is inherent to experiments with largely non-chaotic orbital motion and can only be minimized doing spectral cleaning. This is achieved lowering the trap potential during the filling and by ramping up the magnetic fields just before the beginning of storage cycle. This however requires particular care in the design of superconducting magnets and quench circuitry. The results from this experiment are depicted in fig. \ref{IOFFE_result} and lead to $\tau_n=833^{+74}_{-63}~{\rm s}$ \cite{IOFFE_2}.\\

  \begin{figure}\begin{centering}
   \includegraphics[width=7cm]{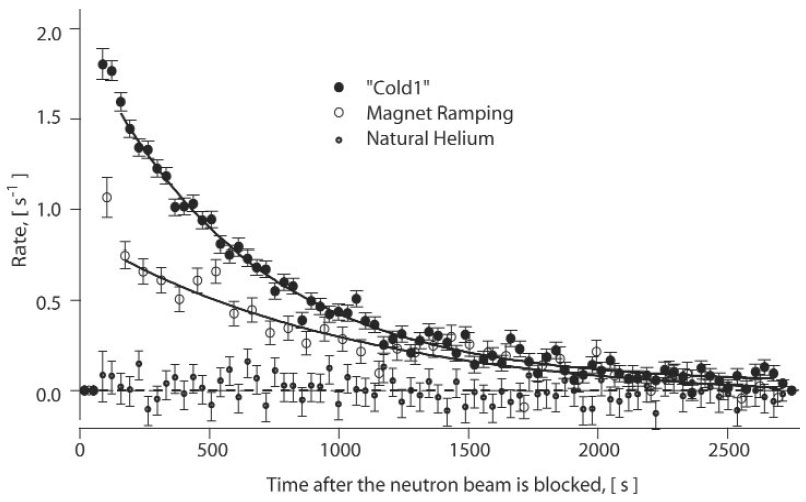}\\
    \caption{Time dependence of the scintillation count rate in the Ioffe trap for three different modes. Steady magnet operation (top curve), ramping mode (central curve) and 'no storage' mode (lowest curve) using natural helium as filling. Denote the change of slope between the two modes of magnet operation \cite{IOFFE_2}.}\label{IOFFE_result}\end{centering}
  \end{figure}
\end{description}
\subsubsection{Material bottles}
Material bottles use highly (neutron-) reflective surfaces which provide wall potentials typically around 200~neV (e.g. beryllium, solid oxygen or fomblin oil). However, such walls are not loss-free due to absorption or up-scattering (inelastic scattering), thus removing neutrons from the storage volume ($\rightarrow \tau_{\mathrm{loss}}$). Therefore, lifetime measurements are performed under different conditions varying the frequency of wall interaction (mean free path $\lambda$) and subsequently extrapolating to $\lambda\rightarrow\infty$. However, care has to be taken that wall conditions are stable with respect to the modifications and that changes in the neutron spectrum during storage are controlled and understood. As discussed above, \textit{marginally trapped} neutrons are an important source of systematic uncertainty.\\

\begin{figure}[h]\begin{centering}
  \includegraphics[width=9.0cm]{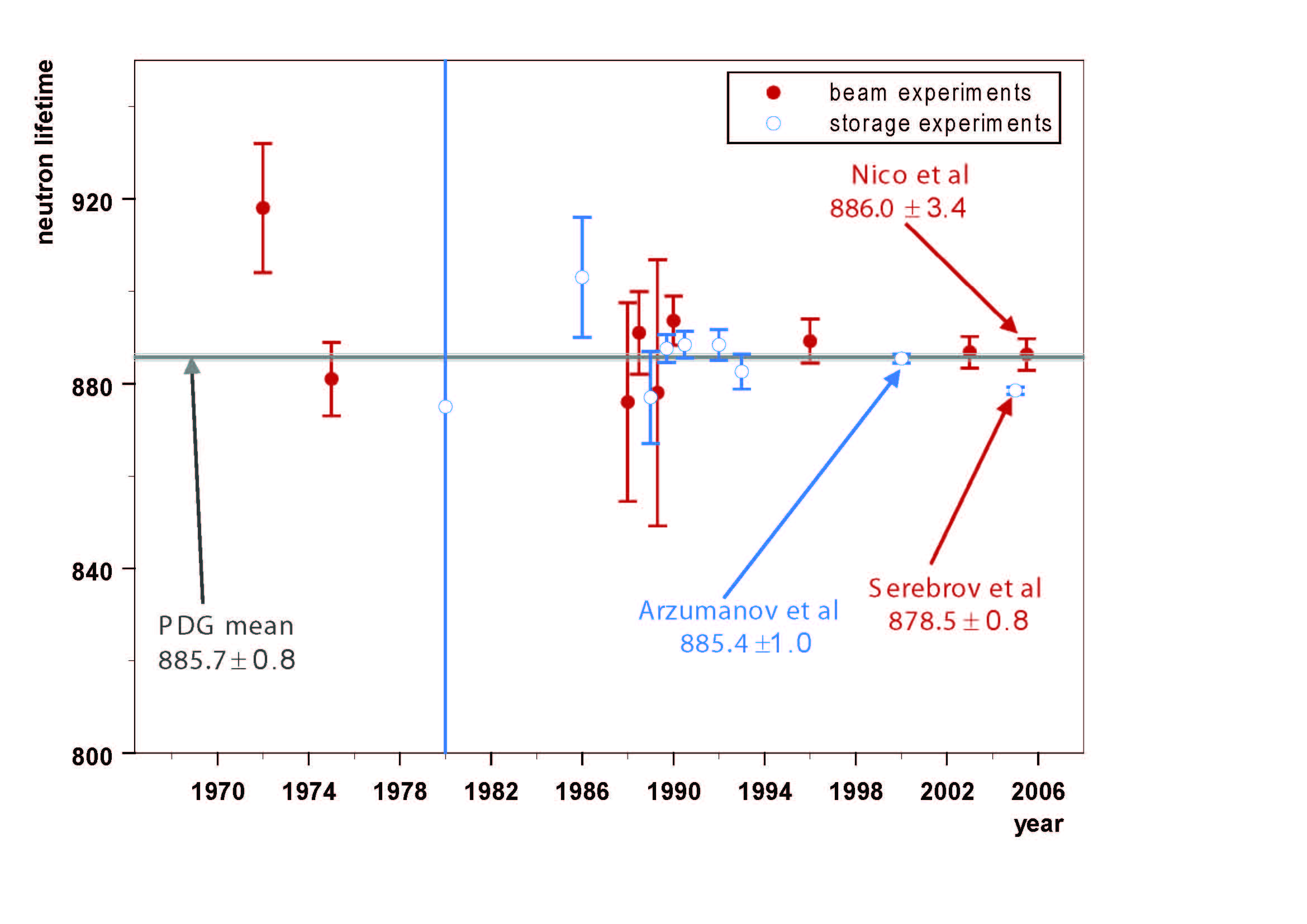}\\
  \caption{Compilation of neutron lifetime measurements sorted by date and method}\label{lifetimes_overview}\end{centering}
\end{figure}

\begin{description}
  \item \textit{Mambo:}  Following the original idea of W. Mampe \cite{Mampe_lifetime} a setup allowing for a temperature variation of the surface coating and the detection of up-scattered neutrons has been used by \cite{Arzumanov}. The setup (fig. \ref{arzumanov_setup}) allowed a precleaning of the spectrum to reduce the effects of \textit{marginally trapped} neutrons, the mechanical change of storage volume as well as a change in wall loss rate by varying the temperature of the fomblin $T\epsilon [-26^\circ{\rm C}, 20^\circ{\rm C}]$.\\
      \begin{figure}\begin{centering}
         \includegraphics[width=8cm]{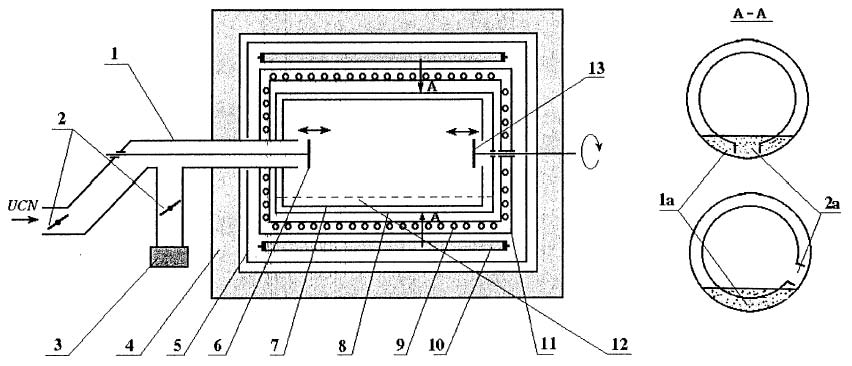}\\
         \caption{Setup of the fomblin coated storage experiment operated at ILL \cite{Arzumanov}. The storage volume can be changed by opening the piston on the right hand side. Up-scattered neutrons can be detected in the neutrons counters surrounding the storage box.}\label{arzumanov_setup}\end{centering}
      \end{figure}
      Drawbacks of this measurement were that no spectral variation of the neutrons could be performed and lack of statistics excluded a systematic investigation of temperature variation (fig. \ref{arzumanov_result}). All in all there was too little information to constrain the systematic uncertainties which thus seem to be underestimated . The result quoted is $\tau_n=885.4\pm 0.9_{\mathrm{stat}}\pm 0.4_{\mathrm{syst}}~{\rm s}$.\\

      \begin{figure}\begin{centering}

         \includegraphics[width=6cm]{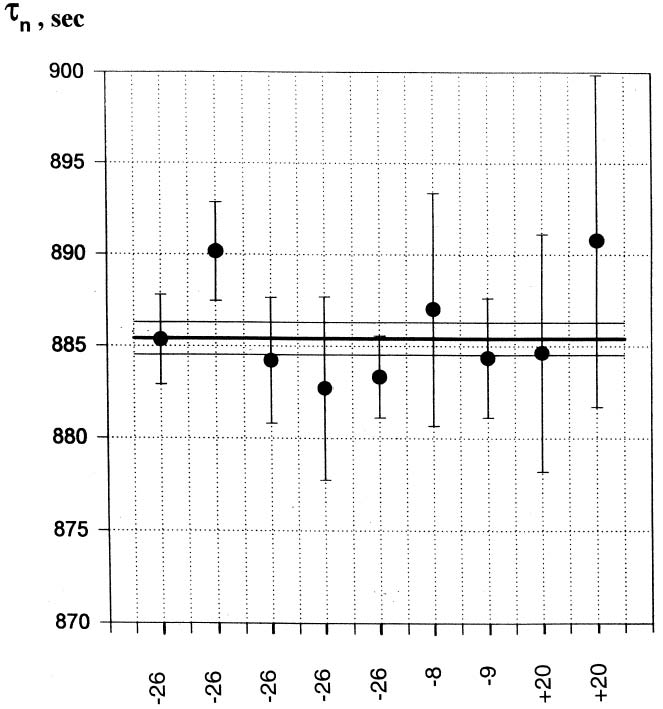}\\
         \caption{Extracted values for $\tau_n$ for different operating temperature of the fomblin coating. Note the large statistical errors not allowing to study wall loss effects at a level below $\pm 2$-$3~{\rm s}$. The data were thus combined and no systematic error was quoted for this study \cite{Arzumanov}.}\label{arzumanov_result}\end{centering}
      \end{figure}

  \item \textit{PNPI experiment:} The newest and possibly most precise storage-type experiment was performed using a gravitationally closed material bottle coated with fomblin oil kept at very low temperature (very small wall loss rate of $\eta=2\cdot 10^{-6}$). The wall loss rates were varied using two different storage volumes of different size (and shape) as well as analyzing the surviving neutrons with respect to their energy. In each case excellent storage time constants were achieved. Quality tests for the coating using an absorptive Ti bottle treated with the same fomblin coating resulted in storage time constants of $\tau_{\mathrm{storage}}\cong 860~{\rm s}$, close to the expected neutron beta lifetime.\\

      \begin{figure}[h]\begin{centering}
        \includegraphics[width=7cm]{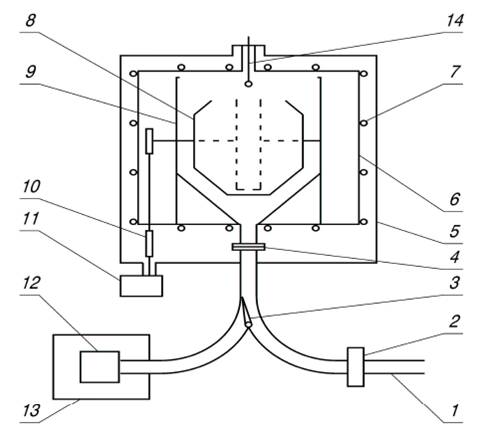}\\
        \caption{Setup of the latest material bottle experiment from PNPI \cite{serebrov_1}. For filling and emptying the bottle can be rotated, thus adjusting the vertical position of the open top.}\label{PNPI_setup}\end{centering}
     \end{figure}

      The stepwise emptying of the storage volume after the storage cycle was performed using the velocity selective gravitational potential such that spectral information was retained (see fig. \ref{PNPI_setup} for the experimental setup). \\

      \begin{figure}[h]\begin{centering}
        \includegraphics[width=6cm]{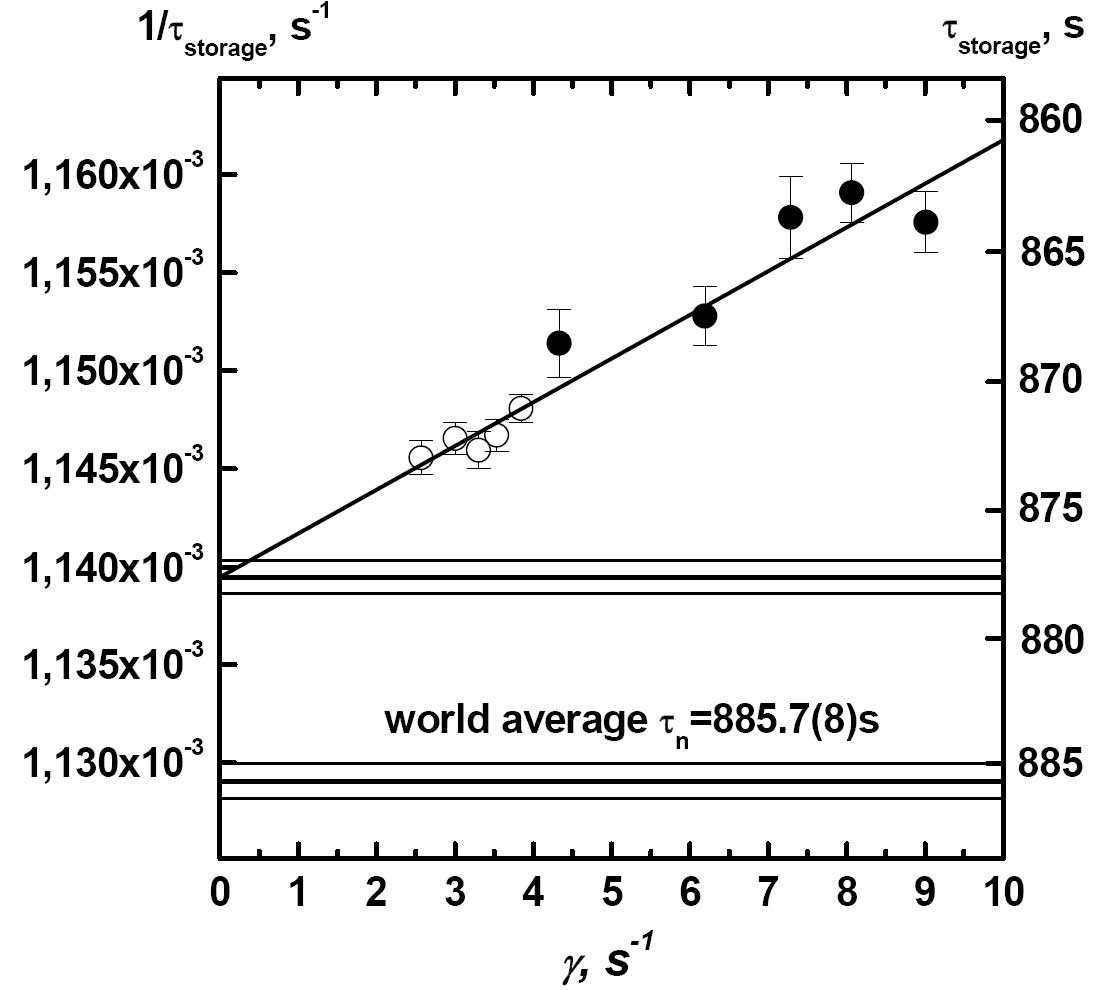}\\
        \caption{Results of the storage cycles with different loss rates using UCN energy selection and two mechanically different storage devices (see the two sets of data points). The data from the two experiments were combined to give the final result for $\tau_n$.}\label{PNPI_lifetme_gamma}\end{centering}
     \end{figure}

      The method of lifetime extraction was based on the calculation of the mean free path $\lambda$ for each measurement and extrapolating to $\lambda^{-1}=0$. Possible corrections were simulated and applied. The only drawbacks of this measurement were the absence of detection of spurious losses of \textit{marginally trapped} neutrons and the statistical limitations which did not allow to investigate independently systematic effects using different setups. Thus, measurements were combined making a determination of a systematic dependence on various parameters impossible (each individual test can at best be constrained to the statistical error of individual measurements). Figure \ref{PNPI_lifetme_gamma} shows the result obtained with different loss rates using two different setups and thus two independent data sets. The data were either combined pairwise with subsequent weighted averaging or combined globally which lead to different values for $\tau_n$ and different statistical errors, thus pointing to some hidden systematics and correlations. At any rate, the global data fit finally used does not allow an independent check on systematics with respect to bottle size or shape with the accuracy quoted by the authors.\\
      The result published is $\tau_n=878.5\pm 0.7_{\mathrm{stat}}\pm 0.3_{\mathrm{syst}}~{\rm s}$ where the systematic error mainly included uncertainties estimated from MC simulations. Although being the experiment with the best inherent storage time constant and thus smallest corrections to the directly measured value the uncertainties quoted seem over optimistic.\\
\end{description}
In summary, the two latest storage experiments (and the most precise ones according to their own uncertainties quoted) differ by 7 seconds or $6~\sigma_{\mathrm{meas}}$, the latter number being smaller if more conservative error treatment is assumed. The overall situation for $\tau_n$ is depicted in fig. \ref{lifetimes_overview}. The dilemma is obvious although a recent statistical analysis of lifetime experiments \cite{Hanson} using Student t-distributions only asks for an overall error scaling of 1.16 to accommodate all measurements.
\section{$\tau_n$ and the CKM matrix}
\begin{figure}[h]\begin{centering}
  \includegraphics[width=7.5cm]{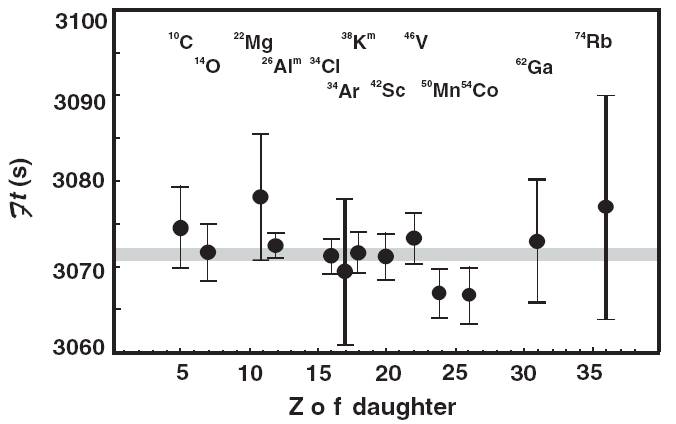}\\
  \caption{Experimental nuclear $\textsl{F}t$-values obtained in ($0^+\rightarrow 0^+$)-transitions from many different nuclei after applying corrections due to nuclear effects. }\label{nuclear_FT_values}\end{centering}
\end{figure}

In eq. \ref{V_A_3} the relation of $\tau_n$ to $V_{ud}$ is depicted. The statistically most precise determination of $V_{ud}$, however, presently comes from nuclear $\beta$-decays ($0^+\rightarrow 0^+$) which involves only $G_{\mathrm{F}}$, radiative corrections ($\delta_{\mathrm{R}}^\backprime$) and various nuclear corrections.

\begin{eqnarray}\label{nuclear_fermi_transitions}
\textsl{F}t=ft(1+\delta_{\mathrm{R}}^\backprime)(1+\delta_{\mathrm{NS}}-\delta_{\mathrm{c}})=\frac{K}{2G_{\mathrm{V}}^2(1+\Delta_{\mathrm{R}}^{\mathrm{V}})}\nonumber \\
|V_{ud}|^2=\frac{K}{2G_{\mathrm{F}}^2(1+\Delta_{\mathrm{R}}^{\mathrm{V}})\overline{\textsl{F}t}}\nonumber
\end{eqnarray}
Many nuclei have been measured and their Q-values determined with high precision. This combined with new calculations for nuclear corrections $\delta_{\mathrm{NS}},\delta_{\mathrm{c}}$) yield a variation for $\Delta\textsl{F}t=2.7\cdot 10^{-4}$ (fig.\ref{nuclear_FT_values}) \cite{hardy08} and a value for $|V_{ud}|=0.97418(26)$. The data show excellent consistency and no obvious remaining dependence on the nucleus used.
\par
For many years the results from neutron decay (lifetime and beta-asymmetry) caused a discussion on the unitarity of the CKM matrix. Based on these results a $3\sigma$ deviation from unitarity was claimed. In light of an alternate value for $\tau_n$ and new progress on other elements of the CKM matrix we shall review this issue again.
\par
\begin{figure}[h]\begin{centering}
  \includegraphics[width=7.5cm]{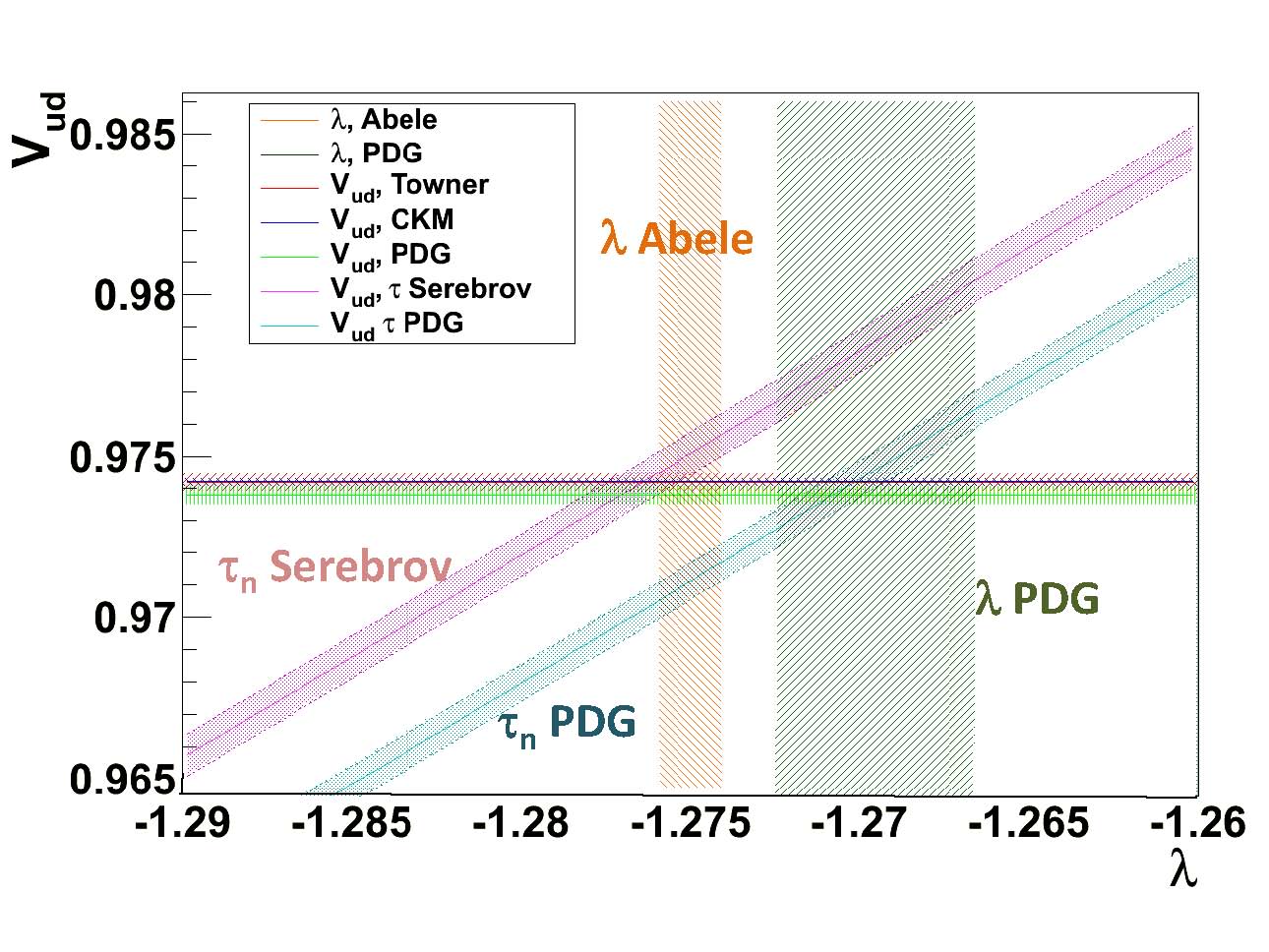}\\
  \includegraphics[width=7.9cm]{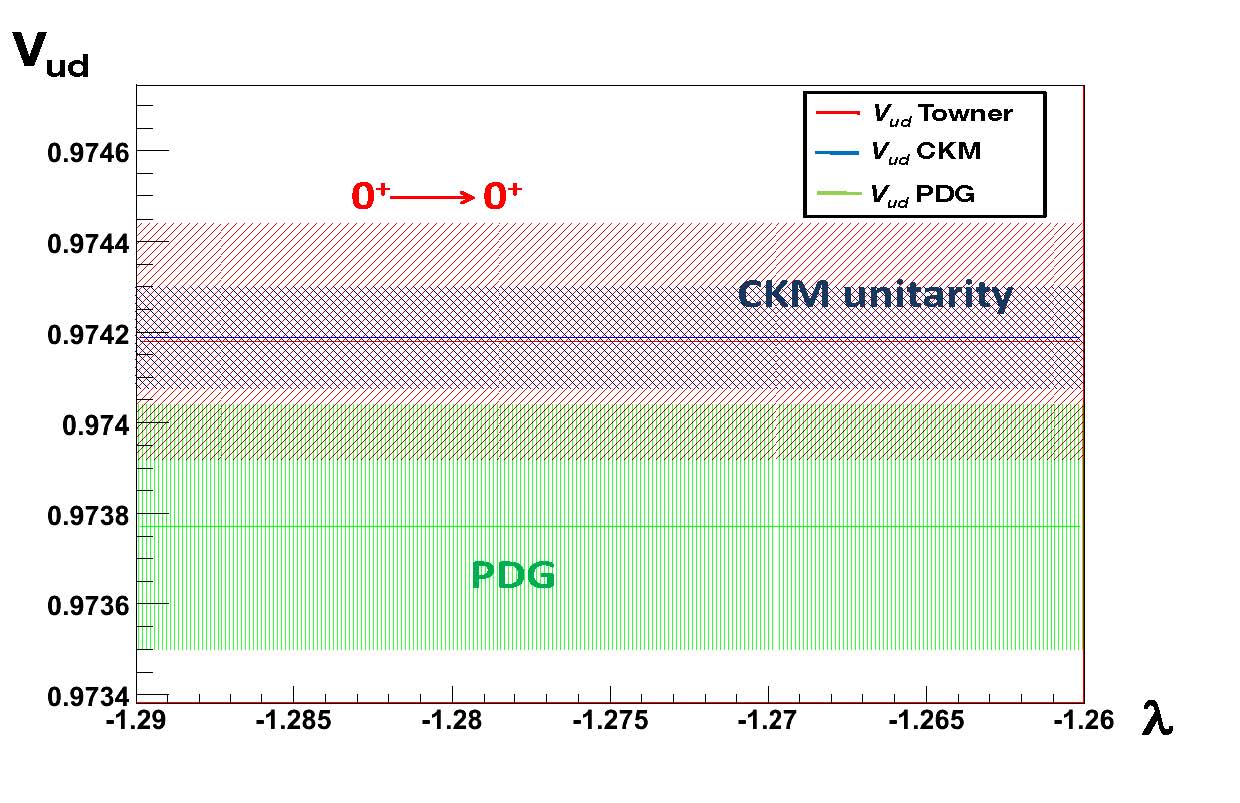}\\
  \caption{Dependence of $V_{ud}$ on the neutron coupling constants $\lambda$ for various input hypotheses.}\label{unitarity}\end{centering}
\end{figure}

Semi-leptonic kaon decays have been revisited in the last years both, from the side of experiments and theoretically concerning the calculation of form factors. These efforts yield a new value for $|V_{us}|=0.2246(12)$ (from \cite{Moulsen}). Using nuclear and kaon decay data we conclude excellent agreement of the CKM matrix with the unitarity hypothesis to a level of $2\cdot 10^{-4}$.
This is also demonstrated in fig. \ref{unitarity}.

In turn we might ask about the value expected for $\tau_n$ as inferred from other experiments. We recall:
\begin{equation}\label{tau_from_vud}
     \tau_n=\frac{1}{\mid V_{ud}\mid^2}\frac{4908.7\pm1.9~{\rm s}}{(1+3\lambda^2)}
\end{equation}
and arrive at the following table:

\begin{table}[h]\begin{centering}
\begin{tabular}{|c|c|c|}
  \hline
  $V_{ud}$ & $\lambda$ & $\tau_n^{\mathrm{calc}}$ \\ \hline
  Hardy \& Towner \cite{hardy08} & PDG \cite{PDG06} & $886.5\pm 3.45~{\rm s}$ \\
  Hardy \& Towner \cite{hardy08}  & Abele et al. \cite{Abele_lambda} & $880.1\pm 1.30~{\rm s}$ \\
  PDG \cite{PDG06}& PDG \cite{PDG06} & $887.2\pm 3.46~{\rm s}$ \\
  PDG \cite{PDG06}& Abele et al. \cite{Abele_lambda} & $880.9\pm 1.32~{\rm s}$\\
  \hline
\end{tabular}
  \caption{Compilation of expected lifetime values $\tau_n^{\mathrm{calc}}$ assuming different values for
  $V_{ud}$ and $\lambda$ (see eq. \ref{tau_from_vud}).}
  \label{table_tau}
  \end{centering}
\end{table}
\section{Future Experiments}
\label{future_measurements}
Considering the systematic limitations for \textit{in-beam} experiments (e.g. absolute neutron-counting efficiencies) all future activities seem to concentrate on storage experiments. Here in turn magnetic storage is much favored owing to the absence of wall losses. The main task of new experiments must be the understanding and precise study and possible elimination of systematic effects limiting the precision of present measurements. These systematics comprise:
\begin{itemize}
  \item UCN spectrum shaping (energy band selection) to understand possible velocity dependent effects
  \item Detection of loss neutrons (set limits on $\tau_{\mathrm{loss}}$)
  \item Combined technique of neutron counting and real-time detection of decay particles (allowing to obtain precise time information without MC simulation on effective emptying dynamics)
  \item Change of trap depth, volume or shape to study geometrical effects including closed orbit motions\\
\end{itemize}
In addition, future experiments should prepare for blind analysis \cite{blind analysis}, thus performing all possible checks and applying the necessary corrections without the exact knowledge of the final result, which by proper preparation of the data will only be revealed applying a final correction factor randomly chosen beforehand and well hidden.\\
The precision of each systematic effect will be governed by the corresponding statistical accuracy and should thus be of the same order as the final statistical accuracy aimed for. This calls for large trap size and new strong UCN sources currently under construction in many neutron laboratories (e.g. ILL, SNS, FRMII, PSI, Triga-Mainz).
\par
The aim of future experiments is twofold, one being to resolve today's discrepancies for which precisions of about 0.5-1~s are sufficient. The final aim for $\delta\tau_n/\tau_n$ must be an accuracy better than nuclear beta decays ($\delta\tau_n/\tau_n<10^{-4}$) and thus $\delta\tau_n<0.08~{\rm s}$. At this point and assuming an adequate precision in $\lambda$, the determination of $V_{ud}$ will be limited by the knowledge of radiative corrections only.

\subsection{Future magnetic trapping}
Magnetic trapping of UCN is presently followed in two ways, superconducting traps or traps built from permanent magnets. The latter have the advantage of allowing complex magnetic structures and possibly low costs while the former one offer superior possibilities in changing trap parameters.
\subsubsection{Magneto-gravitational trapping}  Ultra-cold neutrons with $|v_n|<5~\rm{m/s}$ can be stored in magnetic multipole fields producing large field gradients. Using a gravitational barrier of about 100~neV/m (in vertical direction) the top lid can be left open \cite{zimmer00}. Care has to be taken to avoid field free regions where spin flip can occur or regions where the adiabaticity condition (time dependent change of the magnetic field strength seen by the neutron small compared to the Larmor spin-precession frequency) is not fulfilled, again leading to spin flips and thus turning low-field seekers (repelled from the magnetic walls) to high-field seekers. In order to monitor such cases the high-field seekers have to be extracted by means of mechanical holes in the magnetic wall. General concern for such devices is filling from an external source (internal source filling has been discussed in section \ref{NIST}).\\

\begin{description}
  \item\textit{Permanent magnet trap:}
The first such trap built from permanent magnets (1~T field strength at the surface) was recently used to store neutrons \cite{eshov1} (see fig. \ref{lifetime_eshov_1}). UCN enter from below the trap through a solenoid which acts as a closed valve when powered (in a later version neutrons were filled from the top by means of an elevator system). The solenoid is ramped down for emptying the bottle and neutrons counted. In order to reflect possible high-field seekers and reduce possible losses at the walls these were covered with fomblin oil. The UCN density obtained was about $0.11~{\rm{cm}}^{-3}$ with a bottle size of 3.6~l. A first value for a neutrons storage time results to $\tau_{\mathrm{storage}}=878.2\pm 1.6~{\rm s}$. However, no systematic studies could be undertaken and thus it constitutes a very promising storage time constant for a follow-up experiment with 20 times increased storage volume. Operated at ILL the statistical accuracy obtained in 50 days will be about 0.5~s.
\begin{figure}[h]\begin{centering}
  \includegraphics[height=6.5cm]{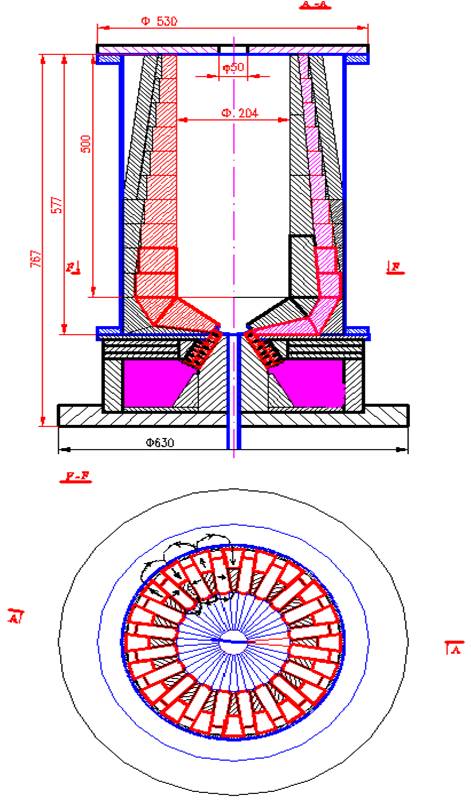}\\
  \includegraphics[width=3.5cm]{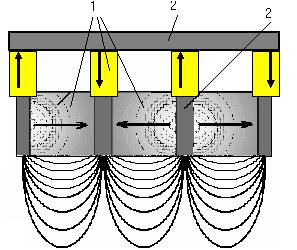}\\
  \caption{Sketch of the magneto-gravitational trap based on permanent magnets. Injections occurs from below through a solenoid. Lower picture: Detailed view of a section of the wall structure with return iron and field lines drawn in \cite{eshov1}.}\label{lifetime_eshov_1}\end{centering}
\end{figure}\\

  \item \textit{PENeLOPE} is the superconducting analog and produces the holding field by a set of cylindrical coils where neighboring coils are operated with opposite current direction, as shown in fig. \ref{penelope_setup} \cite{picker}. Such coils provide the field for the bottom, the outer cylinder wall and an inner cylinder wall which shields the super-conducting race tracks coils, which in turn provide a finite magnetic field perpendicular to the solenoid fields all over the storage volume. The trap walls facing the storage volume are made from an insert covered by an electrically insulating UCN reflective surface. An underlying electrode structure forms the electrical field guiding decay protons upwards, outside of the storage volume, where a proton detector system will be installed \cite{mueller}.\\
  \begin{figure}\begin{centering}
    \includegraphics[height=7.5cm]{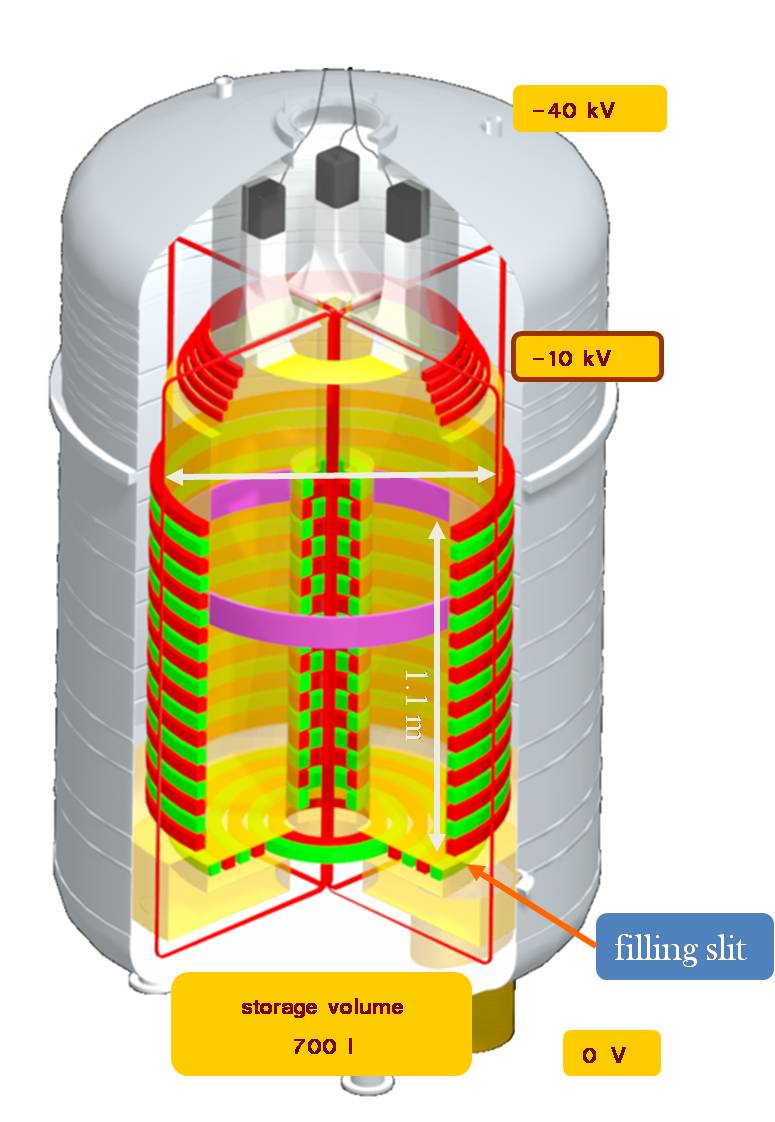}\\
    \caption{Sketch of the experimental setup for the magneto-gravitational trap PENeLOPE (see text). Seen are the coil system (alternating colors), the proton extraction electrodes and the cryo- and vacuum vessels and the absorber ring.}\label{penelope_setup}\end{centering}
  \end{figure}
  The effective magnetic induction at the surface will be about 2~T ($V_{\mathrm{wall}}^{\mathrm{magn.}}=120~{\rm neV}$). The bottle height is about 120~cm, providing a gravitational potential slightly above the wall potential. An absorber ring is installed inside the storage volume to remove neutrons above $v>v_{\mathrm{crit}}$ and thus avoid the problem of marginally trapped neutrons. In addition, the field configuration provides mostly chaotic neutron trajectories.\\
  Filling of the trap is performed through slits between the outer cylinder wall and the bottom and requires the $B$-field to be turned off. Fast ramping of the full system ($T_{\mathrm{ramp}}\approx 100~{\rm s}$) is thus mandatory. Spin flipped neutrons will partially escape through the injection slits where they are funneled to a neutron detector. \\
  The proton detector system consists of a large area thin film CsI-scintillation detector read by Large Area Avalanche Photodiodes (LAAPDs) from the side (fig. \ref{penelope_detection}) and is effectively electron blind.
    \begin{figure}\begin{centering}
    \includegraphics[width=7.5cm]{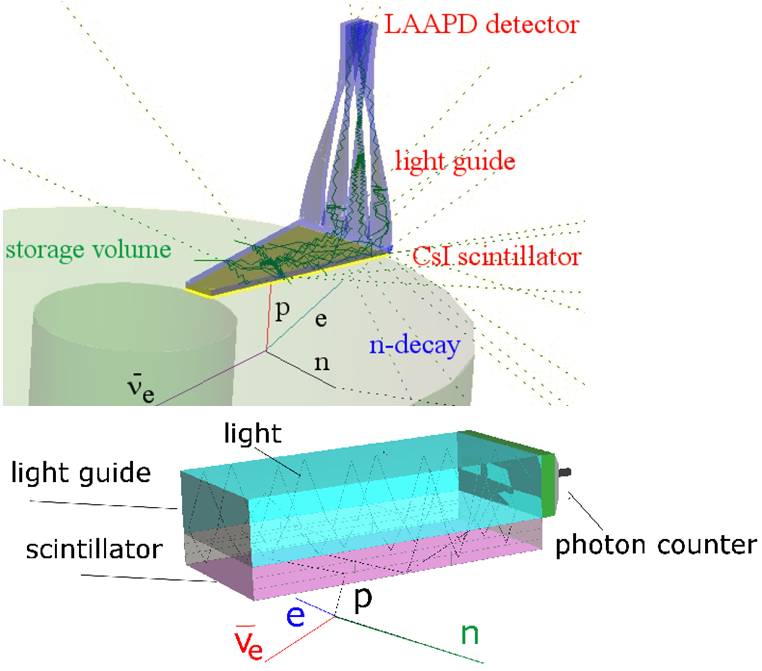}\\
    \caption{Sketch of the proton detection system based on thin film anorganic scintillators with UV LAAPD readout. The detectors are quasi electron blind \cite{mueller}.}\label{penelope_detection}\end{centering}
  \end{figure}
Neutron extraction after a preset holding time will be preceded by a down-ramping of the coil system and neutrons are counted in the same system as used for the detection of possible spin-flip neutrons during storage.\\
The volume of the trap is about 700~l, allowing statistical precisions of about 0.1~s in a few days only. With all installations this experiment provides the largest number of checks for effects possibly modifying the pure neutron $\beta$-decay lifetime spectrum with very high precision. $\delta\tau\approx 0.1$~s should be reached both, on statistical and systematic grounds.\\
A blind analysis will be performed by using a random variation of the clock signal for the sampling ADC of the proton detector which will only be corrected once the analysis is finalized.
\end{description}

\subsubsection{Ioffe traps and others}
The group at NIST is currently improving their experimental setup based on a Ioffe trap using new large size superconducting coils and an upgraded control system to ramp the large magnetic fields provided by them \cite{yang}.
\par
\begin{figure}\begin{centering}
  \includegraphics[width=7cm]{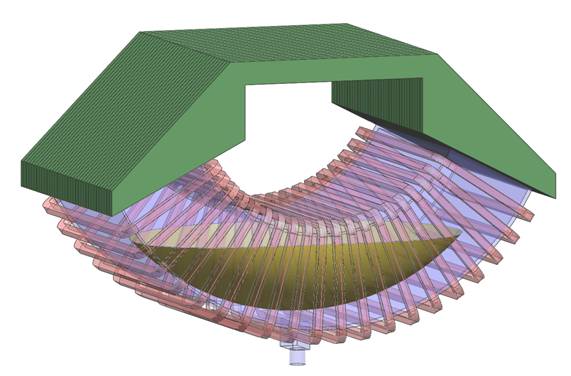}\\
  \includegraphics[width=7cm]{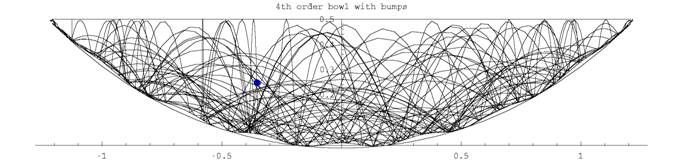}\\
  \caption{Sketch of the setup providing random field orientations for chaotic neutron orbits and a Halbach structure for global neutron trapping. Lower picture: calculated neutron trajectories \cite{bowman}.}\label{ANL_setup}\end{centering}
\end{figure}

The most unconventional setup has been proposed by Bowman at ANL (see fig. \ref{ANL_setup}) \cite{bowman}. The bottom of the trap is made from permanent magnets which provide an almost random field distribution and thus guarantee chaotic neutron orbits (avoiding marginally trapped neutrons). The trap is closed using Halbach-type magnets which also guarantee the absence of field free regions. The trap volume is rather small (4~l) and little is known about studies of systematic effects.
\par
Recently a new scheme has been proposed closely linking UCN production and storage using a novel extraction scheme for UCN out of a superfluid helium bath. Magnetic trapping is performed using a Ioffe trap made from permanent magnet Halbach type octopoles with in situ detection of the decay products \cite{Leung2008}

\section{Conclusions}
The neutron lifetime still is a very important quantity, both in the fields of particle and astrophysics. The experimentally determined values for $\tau_n$ show serious internal discrepancies. This might be connected to some subtle and not understood effects in some of the experiments. We point out that presently the statistical data samples are too small to quote a rather complete systematic confidence interval being as small as published by most authors. The disagreement above also concerns the only high precision experiment obtained using the \textit{in-flight} method (which however quotes larger experimental uncertainties). New experiments, mostly connected with magnetic storage, are in preparation at several sites. High precision experiments, competitive in the extraction of particle physics parameters (as compared to nuclear decays), however, require serious efforts and systematic studies (see section \ref{future_measurements}), the proof of efficient storage itself is thus insufficient. New UCN sources, prerequisite for these efforts, are well under way in many laboratories.\\
If similar efforts as described here will be made for measurements in the $\beta$-decay asymmetry \textit{A} the extraction of particle physics parameters can be performed in a self consistent way with neutron data alone, thus avoiding the remaining issues on nuclear structure corrections.
\section{Acknowledgements}
The author would like to thank the organizers for their kind invitation to this very lively and fruitful workshop. I also owe thanks to many colleagues with whom the question of systematic errors and their estimation were discussed. I also apologize for not having been complete in the presentation of all new ideas and experimental efforts presented in the past or on the occasion of this workshop.




\end{document}